\documentclass[prl,twocolumn]{revtex4}
\usepackage{amsmath}
\usepackage{graphicx}
\begin{document}
\title{Classical density-functional theory for water}
\author{Sahak~A.~Petrosyan, David~Roundy, Jean-Fran\c{c}ois~Bri\`ere and Tom\'as~A.~Arias\\
  {\em Department of Physics, Cornell University, Ithaca, NY 14853}}

\begin{abstract}
We introduce a new computationally efficient and accurate classical
density-functional theory for water and apply it to hydration of
hard spheres and inert gas atoms. We find good agreement with
molecular dynamics simulations for the hydration of hard spheres and
promising agreement for the solvation of inert gas atoms in water.
Finally, we explore the importance of the orientational ambiguity in
state-of-the-art continuum theories of water, which are based on the
molecular density only.
\end{abstract}

\date{in press}
\maketitle
\section{Introduction}
Water is the most abundant liquid on earth and is essential for life.
Water is a called a {\it universal solvent} for its ability to dissolve
both acids and bases, as well as polar and non-polar molecules.

Despite its importance to a wide range of problems in chemistry and
biology, it is still an open challenge to build a first
principles theory of the microscopic structure of water and its
interaction with solutes\cite{Hansen}.  The underlying reason for this is 
the complex interplay between the kinetic energy of molecules and the
 the orientation dependent potential energy of inter-molecule interactions.

For the study of liquids and their interactions with external systems,
two general classes of theoretical methods have been developed.  The
first treats the liquid as a collection of molecules treated either
{\em ab initio} within density-functional theory or with a model
interatomic potential, such as the simple point charge (SPC) model
\cite{SPC} or the 4-site transferable intermolecular potential TIP4P
\cite{TIP4P},
and then uses molecular dynamics (MD) or Monte Carlo numerical methods to
perform statistical averaging over the thermal phase
space\cite{Kollman, Torrie, Carter}.  These methods are intuitively
simple to understand and also relatively straightforward to implement
numerically.  However, because they involve statistical averaging of
many molecules over an exponentially growing configuration phase
space, these methods are numerically very demanding.  The second class
of methods treats the liquid as a continuous media\cite{Ashcroft,
Tomasi, Roux, Sun,Thompson}.  Without the need for thermal averaging
or to represent molecules individually, these latter methods are much more
efficient computationally.  However, such models are generally built
in an empirical way and, generally, there is no systematic way to
improve them.  Nonetheless, significant progress has been made in
understanding the interaction of water with external environments
using this approach, such as the work of Pratt and Chandler~\cite{PC}
on the theory of the hydrophobic effect and the Lum, Chandler, Weeks (LCW)
theory of hydrophobicity\cite{LCW}.

In this work, we explore the nature of the interaction of water with
external environments using a somewhat different approach than that of
Chandler and coworkers and work in a classical density-functional
theory framework.  There are two main advantages of working in this
framework.  First, the classical density-functional theory of liquids
is a continuous theory of the liquid state which is exact in
principle.  Moreover, this framework gives the free energy and the
density profile of the liquid in {\em any} external potential
$V(\vec{r})$ in terms of a single density-functional\cite{Hansen}, so
that study of the hydrophobic effect (liquid in contact with either an
impenetrable wall or an impenetrable hard sphere) and of the
interaction of the liquid with any solute can be carried out in a
single, unified framework.  A number of approximate
density-functionals have been developed for water and applied to the
hydrophobic effect\cite{Sun, Meister}.  Much more demanding theories
which go beyond the average molecular density to consider explicitly
the distribution of molecular orientations in water have also been
developed\cite{Ramirez, Ding}.  All of these density
functional theories, however, prove to be either quite computationally
demanding or provide an over-simplified description.

We begin this work by introducing a new, computationally efficient
density-functional theory for water which accurately reproduces the
hydrophobic effect near hard boundaries.  We then present the first
application of a classical density-functional theory to realistic {\em
ab initio potential energy} surfaces of solutes, applying our theory
to the solvation of the inert gas sequence.  This latter study allows
us to address directly the question of whether explicit orientation
dependence, with all of the concomitant computational demands, is
necessary to provide an accurate description of the solvation of even
the simplest solutes.

\section{Classical density-functional theory}
Classical density-functional theory for liquids follows from a variational
principle which was first established by Hohenberg and Kohn\cite{HK} for
the inhomogeneous electron gas at zero temperature and later generalized to
finite temperature by Mermin\cite{Mermin}.

The theorem states that the equilibrium grand potential $\Omega_0$ of a liquid
at chemical potential $\mu$ in any external potential $V(\vec{r})$ can be found by minimizing,
over all possible thermally averaged molecular densities
$\rho(\vec{r})$, the sum of a {\em universal} functional and the
interaction of that density with the potential,
\begin{equation} \label{eq:cdft}
      \Omega_0=\min_{\rho(\vec{r})}{\left\{F[\rho(\vec{r})]+\int \left(V(\vec{r})-\mu\right) \rho(\vec{r}) \, d\vec{r}\right\}},
\end{equation}
with the density profile $\rho(\vec{r})$ which minimizes the
free-energy giving the exact average density at thermal equilibrium.
$F[\rho(\vec{r})]$ is said to be a universal functional of density
$\rho(\vec{r})$ in the sense that it does not depend {\em whatsoever} on the
external potential $V(\vec{r})$ but only upon the nature of the
solvent itself.  This yields the significant advantage that once a
good approximation is found for the universal functional
$F[\rho(\vec{r})]$, it can be used to study 
the interaction with any possible external potential $V(\vec{r})$,
such as that representing one solute or another.

Even though $F[\rho(\vec{r})]$ is exact in principle, its analytical
form is not known and approximations must be developed.  Generally,
one begins by separating the functional as
\begin{equation}
F[\rho(\vec{r})] = F_{id}[\rho(\vec{r})] + F_{ex}[\rho(\vec{r})],
\end{equation}
where $F_{id}[\rho(\vec{r})]$ is the free energy of ideal gas
whose analytic form is known {\em exactly},
\begin{equation} \label{eq:ideal}
  F_{id}[\rho(\vec{r})]=kT\int\rho(\vec{r})\ln(\rho(\vec{r}))\,d\vec{r},
\end{equation}
and where $F_{ex}[\rho(\vec{r})]$ is defined as the excess free energy
beyond that of the ideal gas and contains all the other terms
responsible for correlations among the molecules.

Our density-functional theory is inspired by a class of
weighted-density functionals\cite{Ashcroft}.  Our method, however, is
computationally much simpler in that it does not require
computationally demanding self-consistent calculations of the weighted
density.  On the other hand, our form does allow us to incorporate much of
the same physics as \cite{Ashcroft} and thus find a form which is
competitive computationally but more
accurate than other, more simplified functionals\cite{Sun}.

\section{Construction of classical density-functional theory for water}

The form we choose to consider is 
\begin{eqnarray} 
  F[\rho(\vec{r})] & = & F_{id}[\rho(\vec{r})] + \int
  f_{ex}(\bar\rho(\vec{r})) d\vec{r} \label{eq:wdft} \\
&& + \iint \rho(\vec{r})
  \chi(\vec{r}-\vec{r}') \rho(\vec{r}')d\vec{r}d\vec{r}' \nonumber
\end{eqnarray}
where $F_{id}[\rho(\vec{r})]$ is given by Eq.~(\ref{eq:ideal}) and
$f_{ex}(\bar\rho(\vec{r}))$ is a local function of $\bar\rho(\vec{r})$,
a Gaussian smoothed density,
\begin{equation} \label{eq:gaussconv}
\bar\rho(\vec{r}) \equiv (2\lambda^2\pi)^{-\frac{3}{2}}\int
    \rho(\vec{r}') e^\frac{-|\vec{r}-\vec{r}'|^2}{2\lambda^2} d\vec{r}',
\end{equation}
and $\chi(\vec{r})$ is an interaction kernel which integrates to zero.
The first term in Eq.~(\ref{eq:wdft}), which is known exactly, ensures
that the short length-scale properties (to which $F_{ex}$ does not
couple) are treated properly.  The second term in Eq.~(\ref{eq:wdft})
ensures that a large-scale phase separation between vapor and liquid
is treated correctly and, often, can capture exactly the experimental
two-particle correlation functions in the bulk liquid\cite{Ashcroft}
for an appropriate convolution kernel in Eq.~(\ref{eq:gaussconv}).
However, we find that, for the experimentally observed two-particle
correlation function in liquid water, the equation for the kernel has
no real solutions.  The solutions have small imaginary parts and thus
cannot be used directly.  The real parts of the solution, however,
closely resemble a Gaussian.  Thus, we employ a Gaussian in
Eq.~(\ref{eq:gaussconv}) and add the last term in Eq.~(\ref{eq:wdft}) to
ensure that the experimental two-particle correlation function in the
bulk liquid is reproduced.

\begin{figure} \centering
\includegraphics[width=8.8cm]{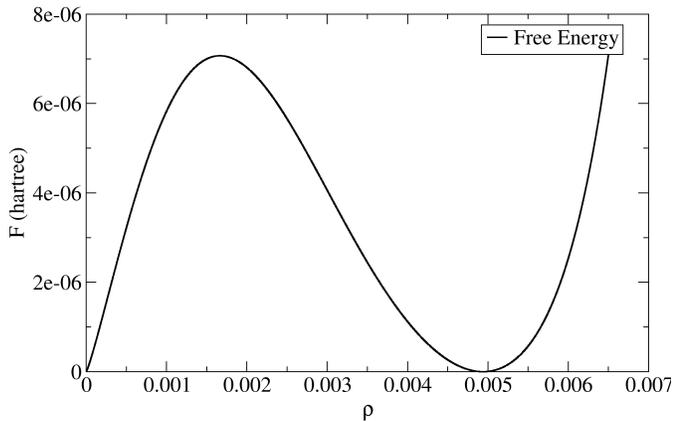}
\caption{Free energy of water as a function of its density at SATP}
\label{fig:free_energy}
\end{figure}

To construct the functions and parameters needed in
Eq.~(\ref{eq:wdft}), we begin by noting that, for the uniform liquid,
the last term gives zero ($\chi$ has zero integral) and
$\bar\rho(\vec{r})=\rho(\vec{r})$, so that $f_{ex}(\rho)$ can be
constructed to reproduce the properties of the uniform liquid exactly.
In practice, we parameterized $f_{ex}(\rho)$ as a sixth-order
polynomial
\begin{equation}
  f_{ex}(\rho) = a \rho^6 + b \rho^5 + c \rho^4 + d \rho^3 + e \rho^2
  +f \rho + g.
\end{equation}
with parameters chosen to reproduce, for bulk water at standard
ambient temperature and pressure (SATP), the density and bulk modulus
of both the liquid and vapor (four parameters), the derivative of the
bulk modulus of the liquid with respect to pressure $\partial
B/\partial P$, and phase coexistence of the liquid and vapor (equality of
grand potential).  Table~\ref{table:parameters}
gives both the fit data and the resulting parameters, and
Figure~\ref{fig:free_energy} presents the final function $F(\rho)$.
Note that in this work we employ mostly atomic units (a.u.) so that Planck's
constant, the electron mass and the fundamental charge all have
numerical value unity ($\hbar=m_e=e=1$), implying that energies are
measured in units of 1~hartree=27.21~eV and distances in units of 1~bohr=0.5291~\AA.

\begin{table}
  \begin{tabular}{|c|c|}
    \hline
    Vapor density&$0.023~kg/m^3$\\\hline
    Water density&$997.1~kg/m^3$\\\hline
    Bulk modulus&$2.187~GPa$\\\hline
    $\partial B/\partial P$&$5.83$\\\hline
  \end{tabular}\\
  \begin{tabular}{|c|c|}
    \hline
    a&$6.630\times10^{9}$\\\hline
    b&$-1.277\times10^{8}$\\\hline
    c&$9.200\times10^{5}$\\\hline
    d&$-2.602\times10^{3}$\\\hline
  \end{tabular}
\begin{tabular}{|c|c|}
    \hline
    e&$8.906\times10^{-4}$\\\hline
    f&$-1.415\times10^{-2}$\\\hline
    g&$1.077\times10^{-10}$\\\hline
  \end{tabular}
  \caption{Properties of uniform bulk water at Standard Ambient Temperature and Pressure (SATP: T=25 $^{\circ}$C, P=100.00 kPa)}
\label{table:parameters}
\end{table}

To construct $\chi(\vec{r})$, we work in the Fourier representation.
As noted above, $\chi(\vec{r})$ integrates to zero so that $\hat
\chi(\vec{k}=\vec{0})=0$.  For $\vec{k}\ne \vec{0}$, we fit to the
experimental pair distribution function $g(r)$ of bulk uniform water
at SATP\cite{ck}.  Finally, the Gaussian width parameter $\lambda =
0.4415$~a.u. has been fit to the surface tension of bulk water
70~mN/m or $4.5\times10^{-5}$~a.u. We emphasize that in constructing
this functional we used only the macroscopic properties of water.

\section{Application to hydration of hard spheres}
To test the accuracy of this approximation we use it to calculate the
free energy of a spherical cavity in water, a standard test case used
in the literature~\cite{LCW,Sun,Huang,Hummer} to explore density
variations in water over all possible length scales. The interaction
potential with a hard sphere is over-idealized because it depends only
on the distance of the water molecule to the hard sphere and does not
depend on the orientation. For instance, for a real water molecule it
is unclear even what point to take to represent the location of the
molecule. The position of the oxygen nucleus is most often taken.

\begin{figure} \centering
\includegraphics[angle=270,width=8.8cm]{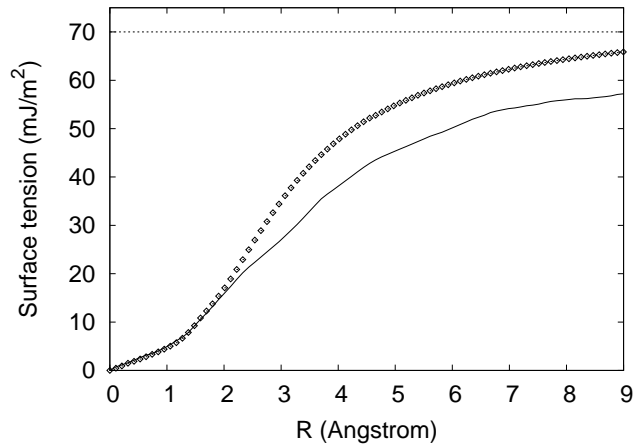}
\caption{Surface tension of a hard sphere in water at SATP. Solid line denotes the results of our classical density-functional theory calculations. Molecular simulation results for a cavity in SPC/E water are indicated by diamonds~\cite{Wolde}. Dashed line corresponds to macroscopic surface tension of water.}
\label{st4}
\end{figure}

Figure~\ref{st4} compares our results for the surface tension (free
energy change per unit area) of a spherical cavity with results of
molecular simulations for SPC/E water\cite{Wolde}.  The figure
verifies that the surface tension approaches in the macroscopic value
in the limit of large radii and show the, for smaller radii, the
surface tension has a strong dependence on radius.  Our results are in
agreement with Lum-Chandler-Weeks\cite{LCW} theory.  The good
agreement with explicit molecular dynamics simulations suggests that
this model gives good quantitative description of the hydrophobic
effect, a central problem of theoretical chemistry.

\section{Application to hydration of inert gas atoms}
The ultimate motivation for inhomogeneous continuous theories of bulk
water is to understand solvation of real solutes, not artificial hard
boundaries.  Within the density-functional framework of
Eq.~(\ref{eq:cdft}), one can simply and rigorously incorporate the
effects of any external potential $V(\vec{r})$ acting on the liquid,
making treatment of real solutes simple in principle, provided a
static potential $V(\vec{r})$ accurately describes the interaction of
water with the solute. For our next test, we treat a simple but
challenging problem for which there is experimental data, the solvation
of inert gas atoms.

The experimental
solubilities of the inert gasses in water
give the corresponding solvation free energies directly.
In the dilute limit, the
free energy of solvation $\Delta \Omega$ relates to the mole fraction
solubility through the relation
\begin{equation}
\Delta \Omega = k_B T \log\left( \frac{n_{\text{gas}}}{X_1 n_w} \right) = k_B T \log\left( \frac{1 \text{atm}}{k_B T X_1 n_w} \right)
\end{equation}
where $X_1$ is the mole fraction solubility in water and $n_w$ is the
number density of liquid water.

The experimental data, which Table~\ref{table:solvation} summarizes,
show an interesting trend.  Although the solvation energies of hard
spheres increase with size, the solvation energies of inert gas atoms
tend to {\em decrease}.  (Note that for helium, the worst possible
case, we estimate quantum zero-point effects to be quite small, on the
order of 0.005~eV, so that this must be purely an effect of the interactions.)  Thus, a simple cavity model for solvation of inert
gas atoms is not a good approximation and one must use a more realistic
interaction potential $V(\vec{r})$.

\begin{figure} \centering
\includegraphics[width=8.8cm]{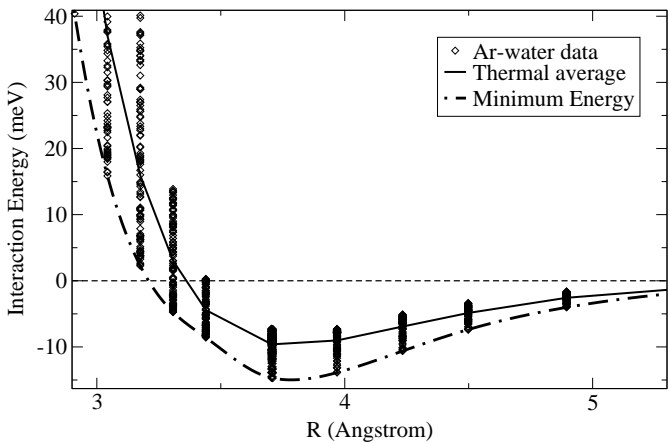}
\caption{Interaction of Argon and a water molecule. Each diamond correspond
to a specific orientation. Also presented are the Boltzmann average potential
(solid line) and the minimum potential (dot-dashed line).}
\label{fig:int}
\end{figure}

We determined the potential energy of interaction between a water
molecule and an atom of each of the inert gasses in
Table~\ref{table:solvation} directly through {\em ab initio}
density-functional theory calculations within the generalized gradient
approximation (GGA) \cite{GGA}.  For these calculations we take
$\vec{r}$ to be the displacement between the nucleus of the inert gas
and the nucleus of the oxygen atom and considered one hundred
different orientations for the water molecule, with these orientations
optimized to sample the angular dependence
evenly according to the procedure of Womersley and
Sloan\cite{Womersley}.  Figure~\ref{fig:int} summarizes the results
for argon, which was typical of all of the inert gasses considered in
this work.

The scatter in the interaction energy at each distance $r$ shows a
strong dependence on the orientation of the water molecule,
particularly at closer distances.  The question then immediately
arises of how to couple to any continuum theory based solely on the
molecular density $N(r)$ with no information about the orientations of
the solute.  As the coupling in such
{\em density-only} functional theories takes the form $\int V(\vec{r})
n(\vec{r}) d\vec{r}$, a choice must be make to to define a unique
value for the potential $V(\vec{r})$ for each point $\vec{r}$. One
approach would be to assume that each water molecule independently
assumes its most favored orientation for a given distance, thereby
defining $V(r)$ as the minimum energy envelop ($V_{min}$) of the data
in Figure~\ref{fig:int}.  Another approach would be to take the
thermal average interaction under the assumption that
the water molecules choose their orientations independently
of each other, thereby defining $V(r)$ as the Boltzmann weighted 
average ($V_{kT}$)  of the data at each distance.

Table~\ref{table:solvation} summarizes the results of the minimum free
energy of our functional for both of these approaches. For helium and neon experimental and theoretical results agree quite well. Also the difference between using $V_{min}$ and $V_{kT}$ is less than $0.04 eV$. As the size of atoms gets bigger the discrepancy also increases suggesting that orientational ordering of water molecules around the atoms is playing an important role and a density-only description is not enough.

\begin{table}
  \begin{tabular}{|c|c|c|c|}
    \hline & Exp. (eV) & V$_{kT}$ (eV) & V$_{min}$ (eV)\\\hline
    Helium &$0.12$ &$0.12$ & $0.09$ \\\hline
    Neon &$0.12$ & $0.11$ & $0.07$ \\\hline
    Argon &$0.09$ & $0.22$ & $0.17$ \\\hline
    Krypton &$0.07$ & $0.25$ & $0.18$ \\\hline
  \end{tabular}
  \caption{Solvation energies of inert gas atoms}\label{table:solvation}
\end{table}

\section{Conclusions and future directions}

This work presents a simple density-functional theory for water which
gives reasonable overall agreement with molecular dynamics simulation
data for the solvation of hard spheres in water, with quite good
agreement for smaller spheres (radius less than 2.5\AA).  However, as
we move to look at simple solutes, such as inert gasses, we find mixed
results.  While we are able to reproduce the counter-intuitive trend
of decreasing energy with increasing size in the inert gas sequence
when going from He to Ne, we fail to do so for the larger atoms.
Thus, although solvation of hard spheres, whose interaction with water
molecules is generally taken to involve only the location of the
oxygen nucleus and thus contains no orientation dependence, may be
well described by density-only theories, such theories do not
necessarily describe well the solvation of even simple solutes such as
inert gases.  It thus appears that treatment of orientation in some
form is needed to attain results approaching chemical accuracy for
realistic solutes. This work was funded by NSF GRANT \#CHE--0113670.

\bibliography{paper}
\end{document}